\documentclass[conference]{IEEEtran}
\ifCLASSINFOpdf
\else
\fi
\usepackage[cmex10]{amsmath}
\ifCLASSOPTIONcompsoc
  \usepackage[caption=false,font=normalsize,labelfont=sf,textfont=sf]{subfig}
\else
  \usepackage[caption=false,font=footnotesize]{subfig}
\fi
\newcommand{\matr}[1]{\mathbf{#1}}

\usepackage{bm}
\usepackage{stfloats}

\usepackage{pgfplots}
\usetikzlibrary{shapes}
\usetikzlibrary{external}
\pgfplotsset{compat=newest}
\newlength\figureheight	
\newlength\figurewidth

\newcommand{\columnplot}{
\setlength\figureheight{0.25\textwidth}
\setlength\figurewidth{0.38\textwidth}
}	

\usepackage[exponent-product = \cdot]{siunitx}
\usepackage{enumitem} 

\begin{document}
%
\title{Statistical Properties and Variations of LOS MIMO Channels at Millimeter Wave Frequencies}

\author{\IEEEauthorblockN{Tim H{\"a}lsig\IEEEauthorrefmark{1}, Darko Cvetkovski\IEEEauthorrefmark{2}, Eckhard Grass\IEEEauthorrefmark{2}\IEEEauthorrefmark{3}, and Berthold Lankl\IEEEauthorrefmark{1}}
\IEEEauthorblockA{\IEEEauthorrefmark{1}Institute for Communications Engineering, Universit{\"a}t der Bundeswehr M{\"u}nchen, Germany\\
\IEEEauthorrefmark{2}Department of Computer Science, Humboldt-Universit{\"a}t zu Berlin, Germany\\
\IEEEauthorrefmark{3}IHP Microelectronics, Frankfurt (Oder), Germany\\
Email: tim.haelsig@unibw.de}
}


%


\maketitle

\begin{abstract}
Measurement results for millimeter wave LOS MIMO systems are presented with a focus on time variation and multipath propagation. Different system setups are used, including $2\times 2$ and $3\times 3$ MIMO, and involving different synchronization procedures and front-ends. Furthermore, different propagation scenarios are evaluated, covering a wide area of applications. The results show that the LOS component carries significantly more power than the NLOS components, and that frequency selectivity from front-ends should be taken into account when designing these high bandwidth systems. Frequency offsets and other phase variations due to transmit and receive oscillator differences are treated as part of the channel and thus, depending on the synchronization setup, the MIMO system exhibits different time variations, particularly in the case of independent local oscillators. It is also observed that these systems experience significant non-trivial long-term variations in terms of amplitude and phase.
\end{abstract}


\let\thefootnote\relax\footnotetext{This work was supported in part by the German Research Foundation (DFG) in the framework of priority program SPP~1655 "Wireless Ultra High Data Rate Communication for Mobile Internet Access". We are indebted to IHP's system design department for providing some of the measurement equipment and assisting during the measurements.}

%
\IEEEpeerreviewmaketitle

\section{Introduction}
Millimeter~wave~(mmWave) wireless communication systems are expected to play a significant role in the improvement of per user data rates in the next few years, due to the wide available bandwidths \cite{Antes2015}. The combination of millimeter wave frequencies with MIMO techniques makes it possible to achieve data rates of several \SI{}{\giga bit\per\second}, especially if the transmission is line~of~sight~(LOS) dominant \cite{Song2017}. Such links occur, for example, in backhauling or satellite transmission scenarios, where the channel can be assumed almost static or deterministic \cite{Bao2015,Storek2015}. However, the time resolution in these studies is not sufficient to determine effects on the symbol or packet level when \SI{}{\giga Bd\per\second} symbol rates are used. 

In practice the transmission channels, including effects due to the front-ends, are rarely fully static. This is, for example, due to the fact that independent oscillators are usually used at the transmitter and receiver side, generating slightly different carrier frequencies and thus a time varying behavior. This time variation can be lowered by suitable synchronization algorithms but some residual variation should always be expected \cite{Meyr1998}. In MIMO systems two different arrangements can be considered. If there is a shared reference frequency among the front-ends on the transmitter and receiver side, respectively, the channel matrix experiences a common time-varying behavior. For cases where independent oscillators are used for the front-ends, each entry of the channel matrix will exhibit a different time-varying behavior. Such a setup may be necessary when the antennas need to be widely spaced as, e.g., in a satellite system. When combinations of the two cases are used, a common variation between the corresponding channel matrix entries will occur. In order to design suitable synchronization and equalization algorithms, it is necessary to characterize the impact of these effects on MIMO systems.

Another effect that has significant impact on the system performance is the presence of multipath components. It is typically assumed that multipath components due to the propagation environment have significantly less power compared to LOS components at millimeter wave frequencies, e.g., at least \SI{10}{\decibel} difference at \SI{28}{\giga\hertz} \cite{Ko2017} and \SI{38}{\giga\hertz} \cite{Rappaport2013}. For a full system design it is also important to consider how the front-ends, i.e., the properties of their internal components, contribute to the frequency selectivity of the complete transmission chain. It has been observed, e.g., in \cite{Antes2015a,Wu2017,Du2017}, that front-ends operating at such high bandwidths generate significant frequency selectivity which needs to be taken into account, if not removed by some form of compensation during manufacturing.

\begin{figure*}[!t]
\centering
\columnplot
\input{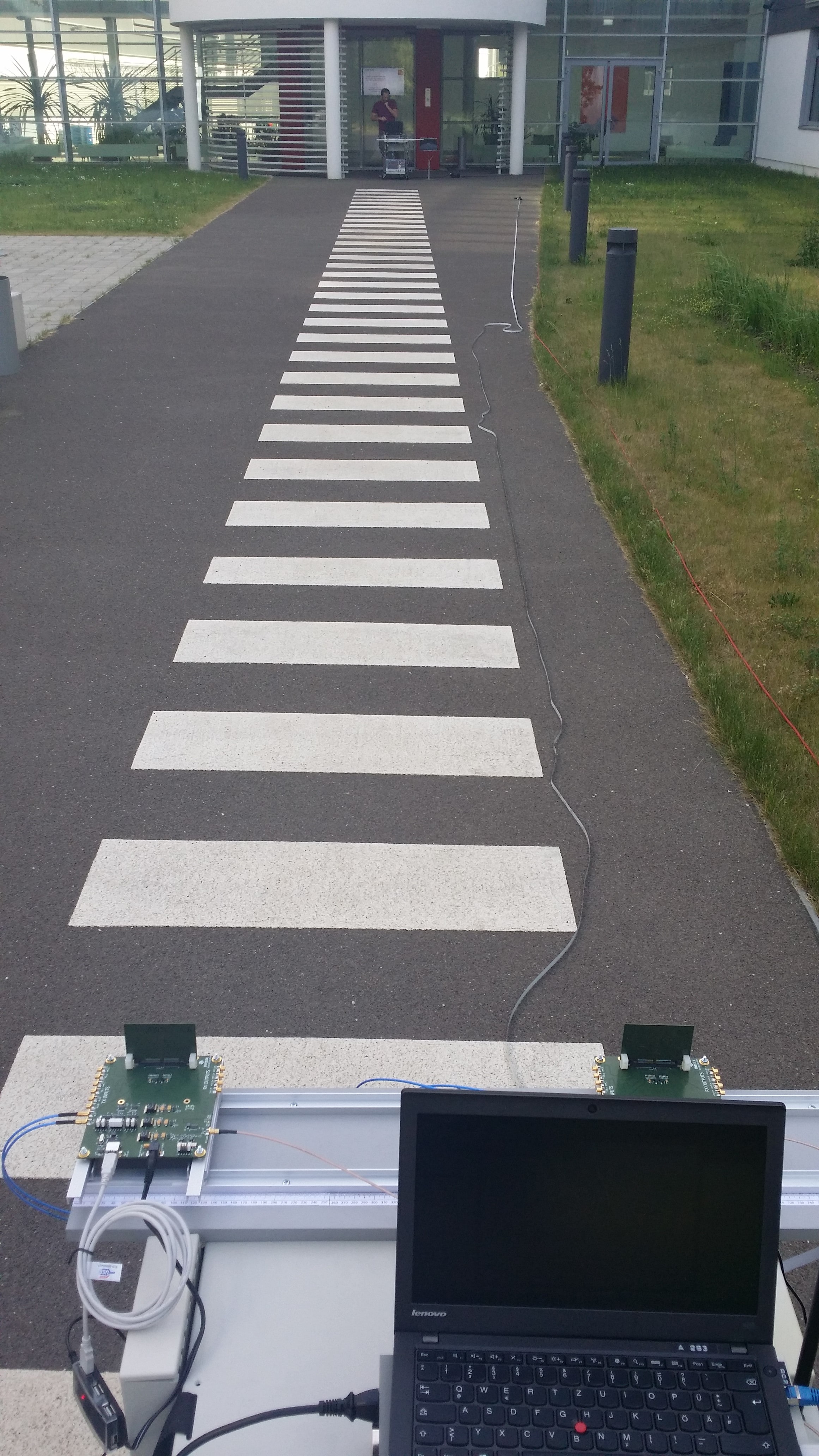}
\caption{Three example measurement scenarios. Left: outdoor $2\times 2$ scenario, FE1, link distance between \SI{10}{\meter} and \SI{60}{\meter}. Center: indoor short-range backhaul $2\times 2$ scenario, FE1, distance of \SI{15}{\meter}. Right: anechoic chamber $3\times 3$ scenario, FE2, distance of \SI{5}{\meter}.}
\label{fig:outside}
\end{figure*}

The following sections of the paper provide measurement results characterizing different behaviors of the LOS millimeter wave MIMO systems with respect to multipath impact and time variations. In particular some statistics and examples for multipath propagation in different MIMO scenarios and environments including the effects of different front-ends at \SI{60}{\giga\hertz} are shown. The paper considers time variations due to frequency differences in the transmit and receive oscillators as part of the channel and investigates their long-term behavior. Furthermore, much shorter time durations, comparable to the symbol or packet level, are considered in order to characterize this time varying behavior of the system, as compared to the existing literature. In comparison to the earlier papers \cite{Halsig2017,Cvetkovski2017}, where it was shown that LOS mmWave MIMO systems exhibit good spatial multiplexing capabilities if designed properly, this paper characterizes the impact of frequency selectivity as well as time variation for these systems.

In this paper, boldface small letters, e.g., $\mathbf{x}$, are used for vectors while boldface capital letters, e.g., $\mathbf{X}$, are used for matrices. Furthermore, $\mathbf{I}_N$ denotes the $N\times N$ identity matrix and $(\cdot)^\dagger$ denotes the pseudoinverse of a matrix.

\section{Measurement Setups, Scenarios, and Channel Estimation}
The measurement data comes from two different campaigns, carried out with different settings and setups as outlined next.

\subsection{Hardware Setup} \label{sec:hardware_setup}

\begin{table}[t]

\renewcommand{\arraystretch}{1.3}
\caption{\SI{60}{\giga\hertz} front-end comparison}
\label{tab:fe_params}
\centering
\begin{tabular}{l|c|c}
 \hline
 & FE1 & FE2 \\
 \hline
 RF bandwidth & \SI{1.8}{\giga\hertz} & \SI{1}{\giga\hertz}\\
 Max. EIRP & \SI{23.5}{\decibel m} & \SI{38}{\decibel m} \\
 Tx/Rx Antenna gain & \SI{7.5}{\decibel i} & \SI{23}{\decibel i} \\
 Noise figure & \SI{7}{\decibel} & \SI{8}{\decibel} \\
 Phase noise @\SI{1}{\mega\hertz} & \SI{-86}{\decibel c\per\hertz} & \SI{-104}{\decibel c\per\hertz} \\
 External ref. clock & yes & no \\
 \hline
\end{tabular}

\end{table}

The general hardware setup consists of two building blocks: the front-ends~(FEs) with antenna and the baseband generating/recording instruments with corresponding processing. Two different FEs were used during the campaigns, an overview of their most important parameters can be found in Table~\ref{tab:fe_params}. The left foto of Fig.~\ref{fig:outside} shows a $2\times 2$ setup with FE1, the right foto shows a $3\times 3$ setup with FE2. With FE1 a synchronization of the carrier frequencies is possible through sharing a \SI{308}{\mega\hertz} reference clock, whereas FE2 does not have an external reference input. In the MIMO measurements there were three clock setups investigated:
\begin{enumerate}[label=(\roman*)]
\item one shared reference clock among all transmitter and receiver front-ends, with FE1,	\label{setup1}
\item one shared reference clock among transmitter \& one shared reference clock among receiver front-ends, with FE1, \label{setup2}
\item independent clock for every front-end, with FE1/2. \label{setup3}
\end{enumerate}

The baseband part consists of arbitrary waveform generators and oscilloscopes with sufficient sampling rates and analog bandwidths to cover the full RF bandwidth of the front-ends. For almost all of the setups a symbol rate of \SI{1.25}{\giga Bd\per\second} was used. Since the sampling clocks of the waveform generators and oscilloscopes were not synchronized, an oversampled signal was captured, and a digital synchronization algorithm including interpolation was used in order to align the clocks in a post-processing step. With approximately aligned sampling, a symbol spaced channel representation suffices and training signals were used in order to determine the variations of the effective transmission channel between the baseband input at the transmitter and baseband output at the receiver.

More detailed descriptions about, e.g., the connection of front-end to baseband, reference clock generation, and sampling synchronization can be found in \cite{Halsig2017,Cvetkovski2017}.


\subsection{Measurement Scenarios}
Different measurement scenarios were employed focusing on two investigation points. First, examining the significance of frequency selectivity in mmWave LOS MIMO systems, either due to multipath components from the environment, or due to distortions in the front-ends. Thus, a variety of short- and mid-range (\SI{5}{\meter}-\SI{60}{\meter}) measurements in different environments, always with a clear LOS path, were carried out. Three of the scenarios can be seen in Fig.~\ref{fig:outside}. While most of the indoor and outdoor scenarios were measured with FE1, both FEs were used in the anechoic chamber in order to determine the frequency-selectivity due to the front-ends, as no multipath from the environment is expected in that case. 

Secondly, the time variation of the channel due to the environment and hardware is investigated. From the scenarios described above, and by changing the synchronization setup as mentioned in the previous section, short- and long-term variability can be deduced.

\subsection{System Model \& Channel Estimation}
In order to estimate the channel responses, consider a frequency selective and time-varying MIMO system to be modeled by
\begin{align}
\mathbf{y}(k) = \mathbf{H}_{L}(k) \mathbf{x}_L(k) + \mathbf{w}(k) \text{,} \label{eq:basicMIMOfo}
\end{align}
where $\mathbf{y}(k)=\begin{bmatrix} y_1(k) & y_2(k) & \cdots & y_M(k) \end{bmatrix}^{\operatorname{T}}$ is the sample vector of the $M$ receive antennas at time $k$, $\mathbf{x}_L(k)=\begin{bmatrix} \mathbf{x}(k) & \mathbf{x}(k-1) & \cdots & \mathbf{x}(k-L+1) \end{bmatrix}^{\operatorname{T}}$ is the space-time stacked transmit vector of the $N$ transmit antennas, with $\mathbf{x}(k)$ being defined similar to $\mathbf{y}(k)$. The parameter $L$ describes the number of multipath components in the system, yielding the time-varying channel matrix $\mathbf{H}_L(k)=\begin{bmatrix} \mathbf{H}_0(k) & \mathbf{H}_1(k) & \cdots & \mathbf{H}_{L-1}(k) \end{bmatrix}$, where each $\mathbf{H}_l(k)$ includes the transfer characteristics between the $N$ transmit and $M$ receive antennas for the $l$th multipath component. $L$ can in general be time-varying but will in this work be fixed, such that it captures the most significant channel effects. Finally, $\mathbf{w}(k)\sim\mathcal{C}\mathcal{N}(0,\mathbf{I}_{M})$ models a spatially and temporally white Gaussian noise process.

For the estimation of the time-varying channel matrix $\mathbf{H}_{L}(k)$, a least-squares approach is used, i.e., 
\begin{align}
\hat{\mathbf{H}}_{L}(k) = \mathbf{Y}_{L_\text{T}}(k) \mathbf{X}_{L,L_\text{T}}^\dagger(k) \text{,} \label{eq:chestimator}
\end{align}
where the training matrix is a block version of the $L_\text{T}$ transmitted training vectors with $\mathbf{X}_{L,L_\text{T}}(k)=\begin{bmatrix} \mathbf{x}_L(k) & \mathbf{x}_L(k-1) & \cdots & \mathbf{x}_L(k-L_\text{T}+1) \end{bmatrix}$. The block of received samples corresponding to the training period is described by the matrix $\mathbf{Y}_{L_\text{T}}(k)=\begin{bmatrix} \mathbf{y}(k) & \mathbf{y}(k-1) & \cdots & \mathbf{y}(k-L_\text{T}+1) \end{bmatrix}$. The channel can then be well estimated if the training matrix fulfills $\mathbf{X}_{L,L_\text{T}}(k)\mathbf{X}_{L,L_\text{T}}^\dagger(k)\approx \mathbf{I}_{N\cdot L}$. In other words, the training sequences possess nearly perfect auto- and crosscorrelation properties. For the measurements pseudo-random sequences that approximately fulfill this property were used. Note that any variations that happen during the transmission of one training block of length $L_\text{T}$ cannot be properly estimated with this method.

Since the attenuation varies significantly due to the different setups, e.g., front-ends, and scenarios, e.g., distances, and because the focus is on the relative behavior between LOS and NLOS, each channel estimate is normalized with respect to the average LOS path power of the corresponding snapshot with
\begin{equation}
\hat{\mathbf{H}}_{L}(k) = \frac{\hat{\mathbf{H}}_{L}(k)}{\frac{1}{MN\cdot K}\sum_{mn} \sum_k\left|\hat{\mathbf{H}}_{0}(k)\right|} \text{.}
\end{equation}

\begin{figure}[!t]
\centering
\columnplot
\input{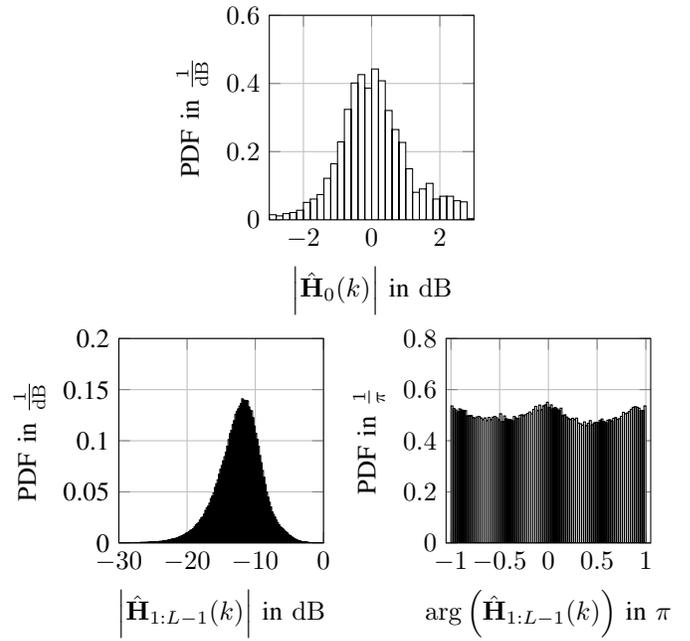}
\caption{Amplitude and phase distributions of 15 different measurement setups, each with 40 recordings, optimally- and ill-conditioned LOS MIMO channel, $2\times 2$ and $3\times 3$, FE1 \& FE2, and frequency offsets removed. Top: amplitude of LOS component. Bottom: amplitude and phase of NLOS components.}
\label{fig:pdfs}
\end{figure}

\begin{figure}[!t]
\centering
\columnplot
\tikzexternalenable
\input{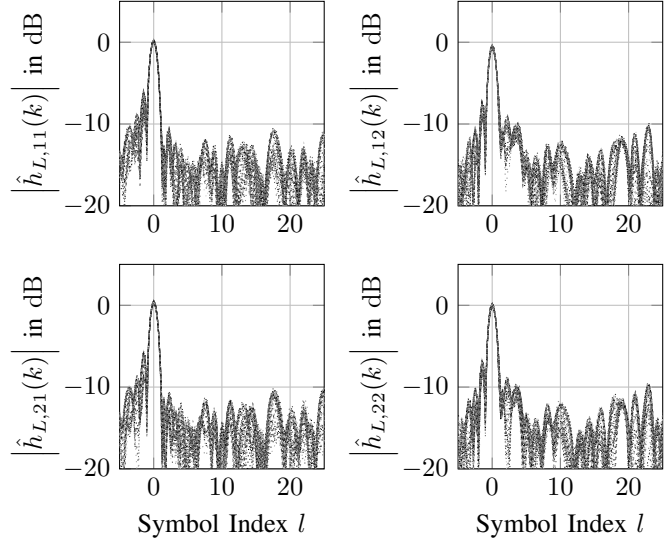}
\tikzexternaldisable
\caption{Superposition of 75 consecutive channel impulse response estimates of an example recording of a $2\times 2$ LOS MIMO setup with FE2 in an anechoic chamber, sample rate \SI{5}{\giga Sa\per\second}, symbol rate \SI{1.25}{\giga Bd\per\second}, distance \SI{5}{\meter}.}
\label{fig:CIRs}
\end{figure}

\section{Measurement Results}
The results are grouped in different categories and presented in the next couple of sections.

\begin{figure*}[!t]
\centering
\columnplot
\subfloat[]{\input{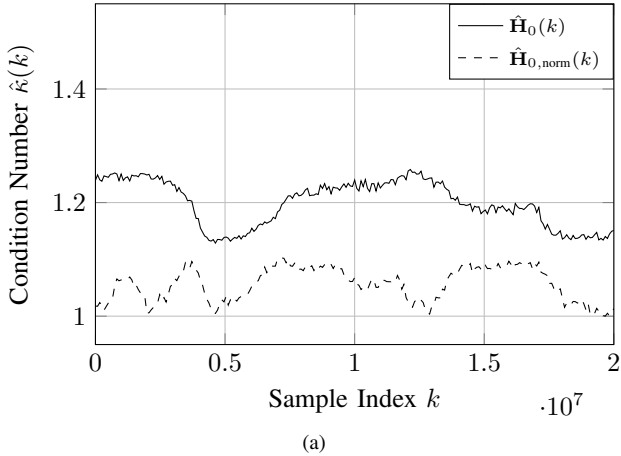}\label{fig:CondNr}}
\hfill
\subfloat[]{\input{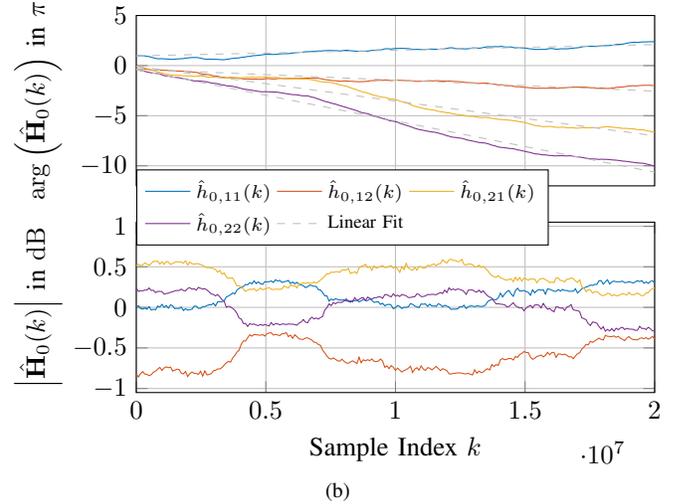}\label{fig:ChannelVar}}
\caption{Example recording of a $2\times 2$ optimally arranged LOS MIMO setup with FE2 in an anechoic chamber, clock setup \ref{setup3}, sample rate of \SI{10}{\giga Sa\per\second} corresponding to a full snapshot length of \SI{2}{\milli\second}: \protect\subref{fig:CondNr}~Condition number variation of the LOS component over time, also including the normalized case where amplitude gain imbalances are neglected; \protect\subref{fig:ChannelVar}~Phase and amplitude variations of the LOS component over time.}
\label{fig:frameVariation}
\end{figure*}

\subsection{Multipath in mmWave LOS MIMO} \label{sec:multipathLOS}
At first the combined statistics for all setups and scenarios are shown in Fig.~\ref{fig:pdfs}. For them, the phase variation due to the frequency offset was removed by estimating it from the LOS component. In total there were five different measurement environments, three of them can be seen in Fig.~\ref{fig:outside}, the other two are short-range indoor scenarios. The estimates of the probability~density~functions~(PDFs) of the amplitude and phase come from 15 different setups, most of them using FE1, each with 40 impulse response snapshots, yielding 600 in total. The setups include both optimally- and ill-conditioned LOS MIMO setups, refer to \cite{Halsig2017}, and $2\times 2$ as well as $3\times 3$. The goal here is to get a rough parameter range that is relevant for mmWave LOS MIMO systems, as in general each environment and setup generates significantly different channels at mmWave carrier frequencies.

The results in Fig.~\ref{fig:pdfs} show a significant difference in power between the LOS $\hat{\mathbf{H}}_{0}$ and the NLOS components $\hat{\mathbf{H}}_{1:L-1}$. While the LOS component has an average relative power of \SI{0}{\decibel}, due to the normalization described above, the NLOS components average relative power is \SI{-11.2}{\decibel}. This result is consistent with others from the literature, e.g., \cite{Ko2017,Rappaport2013}. The variation of the LOS component is due to a couple of factors. For example, with short ranges the antenna pattern has a significant impact, especially for the cross connections, i.e., $m\neq n$. In general this statistic gives a good indication of how the power can vary for the LOS component of mmWave MIMO system setups, when no significant additional effort is spent on antenna alignment. The phases of the NLOS components appear almost uniformly distributed with a tendency to the phase state of the LOS component and its opposite.

\subsection{Short-Term Amplitude Variation and Frequency Selectivity due to Hardware}
In order to investigate the short-term variability and the frequency selectivity due to the front-ends, measurements in an anechoic chamber, where no multipath from the environment is expected, were carried out with FE2 in a $2\times 2$ setup. The amplitude of the channel impulse response estimates for one snapshot are shown in Fig.~\ref{fig:CIRs}. For the plot, 75 consecutive estimates are superimposed, where an estimate was formed roughly every \SI{0.8}{\micro\second}, yielding a complete snapshot length of \SI{60}{\micro\second}. Note that for the symbol rate of \SI{1.25}{\giga Bd\per\second} this corresponds to \SI{75}{\kilo Symbols}.

It is seen that there is only a minor variation in the amplitude of the impulse responses, especially for the most significant components. Furthermore, multiple components about \SI{10}{\decibel} lower than the LOS component can be observed. Since no reflections ought to be coming from the wireless propagation in this environment, they must be due the front-ends, see also \cite{Antes2015a,Wu2017,Du2017}. In fact, by inspecting the plots in the columns of Fig.~\ref{fig:CIRs} it seems that most of the selectivity is coming from the transmitter side, as the column entries have similar significant components at similar positions. Measurements with FE1 showed similar but less severe behavior. This is probably due to the increased RF bandwidth and also due to the fact that the frequency selectivity is harder to measure due to an increased noise floor from phase noise, refer to Table~\ref{tab:fe_params}.

\subsection{Long-Term Variations of the LOS Component}
For checking the long-term behavior of the system, specifically the LOS component, the same setup as in the previous section, i.e., FE2 in an anechoic chamber, is used. The results are presented in Fig.~\ref{fig:frameVariation}, where the condition number of the LOS channel matrix given by
\begin{equation}
\hat\kappa(k) = \frac{\sigma_{\max}\left(\hat{\matr{H}}_0(k)\right)}{\sigma_{\min}\left(\hat{\matr{H}}_0(k)\right)} \text{,}
\end{equation}
with $\sigma_{\max}\left(\cdot\right)$ and $\sigma_{\min}\left(\cdot\right)$ being the largest and smallest singular value of a matrix, respectively, is used as a metric. It has been observed in \cite{Halsig2017} that typical values for properly designed LOS MIMO systems are $1<\kappa \leq 3$. This metric gives direct insight into the spatial multiplexing capabilities of the MIMO system. As was seen in section~\ref{sec:multipathLOS}, and also observed in \cite{Halsig2017}, there is some variation of the amplitude of the channel entries. Thus, the normalized channel matrix $\hat{\matr{H}}_{0,\text{norm}}(k)$, where all paths carry the same amount of power, is also added as a reference.

\begin{figure*}[!t]
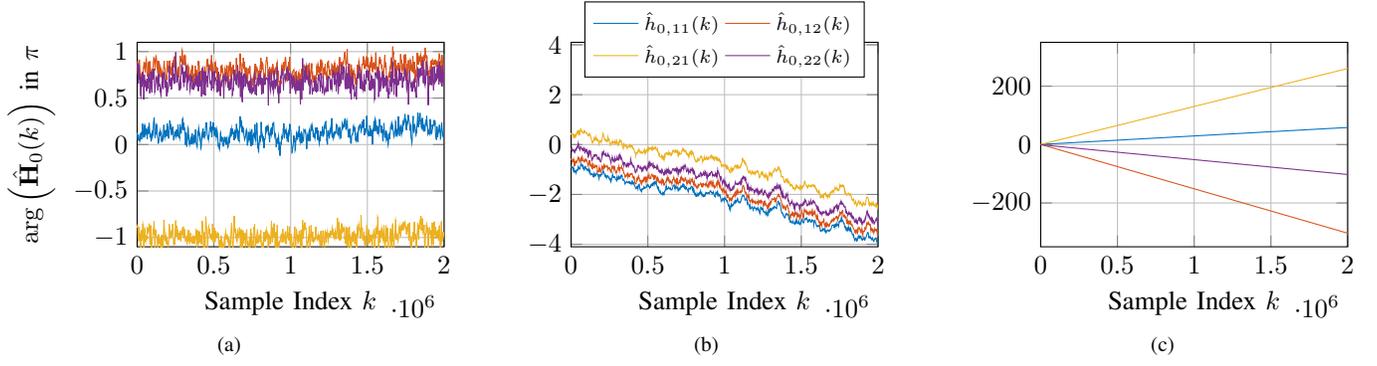

\centering
\columnplot
\subfloat[]{\input{figs/phase_common}\label{fig:phase_common}}
\hfill
\raisebox{-0.0mm}{\subfloat[]{\input{figs/phase_shared}\label{fig:phase_shared}}}
\hfill
\subfloat[]{\input{figs/phase_independent}\label{fig:phase_independent}}
\caption{Example of three different MIMO system synchronization setups in an anechoic chamber, FE1, sample rate of \SI{10}{\giga Sa\per\second}: \protect\subref{fig:phase_common}~Shared clock among all transmit and receive front-ends, setup \ref{setup1}; \protect\subref{fig:phase_shared}~One shared clock on the transmit side and one shared clock on the receive side, setup \ref{setup2}; \protect\subref{fig:phase_independent}~Independent clocks for all front-ends, setup \ref{setup3}.}
\label{fig:clocksetup}
\end{figure*}

The results show that there is a strong correspondence between the conditioning of the channel matrix and the variation in phase and amplitude. Since the setup is static, such variations are suspected to come from the front-ends. In fact, by checking the SNRs of the recordings it is seen that they do not change over time, as also visible in the bottom plot of Fig.~\ref{fig:ChannelVar}. In that plot it is seen that the gains of receivers 1 and 2 change relative to each other, meaning that there is a gain variation in the front-ends over time which needs to be accounted for in a system design. The top plot of that figure shows the phase variation of the channel entries, each of them having a different phase trajectory. This is due to the fact that the frequencies cannot be synchronized with FE2, i.e., only clock setup \ref{setup3} is possible, as will be explored further in the next section. Additionally, the trajectories of the phase are not just simple ramps, compared to the linear fits, but exhibit more complex behaviors, which needs to be taken into account for frequency offset estimation and compensation. Note again that the full length of the recording is \SI{2}{\milli\second} corresponding to \SI{2.5}{\mega Symbols} meaning that these effects are less severe on the symbol level, but nevertheless important for the general system design.

\subsection{Long-Term Phase Variations with different Clock Setups}
In this section the phase variations of the mmWave LOS MIMO channel with respect to different synchronization setups are investigated. For that an anechoic chamber setup with FE1 is used. With FE1 all of the three synchronization setups mentioned in section~\ref{sec:hardware_setup} are possible and the results are shown in Fig.~\ref{fig:clocksetup}. 

Fig. \ref{fig:phase_common} shows the case were a common reference clock is used for all FEs, i.e., setup \ref{setup1}. The phases are relatively constant over the complete recording but have a fixed difference. This difference is partly due to the LOS MIMO channel, which generates phase shifts between the different transmit and receive antennas, and partly due to different initial phases of the carrier frequencies. Although a reference clock is shared, each front-end has an independent phase locked loop which generates the carrier frequency from the reference, and whose phase states are not equivalent. The same can also be observed for the other synchronization setups. In \ref{fig:phase_shared} the phase changes of the system with shared reference on transmitter and receiver side, respectively, are shown, i.e., setup \ref{setup2}. A slowly increasing phase, with approximately the same amount for all entries, can be observed, essentially showing the frequency difference between the transmitter and receiver carrier frequencies generated from the two reference clocks. 
Results for the independent clock setup \ref{setup3} are given in Fig.~\ref{fig:phase_independent}. The phase change for each entry of the channel matrix is different over time and displays the frequency difference of the oscillators of the corresponding transmitter and receiver front-end. The phase change over time is significantly higher compared to the other two cases, as the internal reference clocks have orders of magnitude higher inaccuracies compared to the external ones.

As in the previous section, it can be observed that the phase change is not just linearly increasing but has a more complex behavior, which needs to be taken into account when designing estimation and equalization schemes. For the last case this becomes more obvious when a longer time frame, as in Fig.~\ref{fig:ChannelVar}, is considered.

\begin{figure}[!t]
\centering
\columnplot
\input{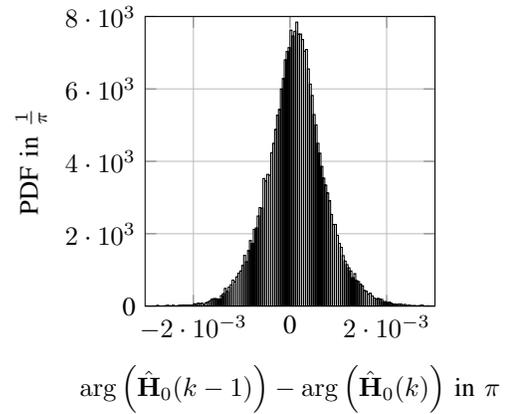}
\caption{Distribution of symbol to symbol phase variations from 15 different measurement setups, each with 40 recordings, optimally- and ill-conditioned LOS MIMO channel, $2\times 2$ and $3\times 3$, FE1 \& FE2.}
\label{fig:phasevar_distribution}
\end{figure}

\subsection{Short-Term Phase Variations on the Symbol-Level}
Finally, the symbol-level behavior of the phase variation is investigated. The complete set of measurements, as in section~\ref{sec:multipathLOS} is used, and the phase difference between two consecutive symbols is estimated.

The distribution of the results can be seen in Fig.~\ref{fig:phasevar_distribution}. The mean value is \SI{1.3e-5}{\pi\per Symbol}. With the often used symbol rate of \SI{1.25}{\giga Bd\per\second} this corresponds, for example, to a mean frequency offset of \SI{8.13}{\kilo\hertz}. As is usually the case, the phase variation on the individual symbol level due to frequency offset is low, but accumulates and needs to be compensated for over time. Note that these plots just give an indication on the average phase change per symbol over time, but lack information about the behavior of the phase in terms of dependence over time.

\section{Conclusion}
This paper provides measurement results for mmWave LOS MIMO systems with different front-ends in different environments, and with different synchronization setups. The analysis shows that in LOS dominant MIMO at \SI{60}{\giga\hertz}, the power level of the NLOS components is at least \SI{10}{\decibel} below the desired LOS component. Although great care was taken when setting up the systems, significant variations of around \SI{2}{\decibel} can still be noted for the LOS component. With automated alignment procedures, the power loss due to the non-ideal alignment could be further reduced. Contrary to the initial expectation, the frequency selectivity due to the front-ends has a dominant impact on the effective channel in comparison to multipath propagation and needs to be adequately compensated.

Amplitude variations over time can be considered insignificant even for a large number of symbols, but are important for the long-term behavior of the system. On the other hand, phase variations are strongly pronounced, particularly for the independent reference clock setup, and exhibit complex behavior that needs to be modeled appropriately for the system design. Additionally, in the independent clock case, the differently evolving phases of each channel matrix entry need to be taken into account for the equalizer design. In other words, frequency offset compensation cannot be done by premultiplying with a common rotation before equalization, but needs to be done for each entry individually. Finally, it should be noted that all the effects discussed are relatively slow compared to the typical symbol durations in millimeter wave systems. To be specific, symbol durations are typically in the order of \SI{}{\nano\second} whereas the changes observed occur in the order of \SI{}{\micro\second}. This means that slowly adapting and tracking receivers may be a very good choice in these systems.

\bibliographystyle{IEEEtran}
\bibliography{references}
%

\end{document}